\documentclass[sigconf]{acmart}

\AtBeginDocument{%
  \providecommand\BibTeX{{%
    \normalfont B\kern-0.5em{\scshape i\kern-0.25em b}\kern-0.8em\TeX}}}

\setcopyright{none}

\usepackage{multirow}
\usepackage{booktabs} 
\usepackage{tabularx}
\usepackage{comment}
\usepackage{sidecap}
\usepackage{subfigure}
\usepackage{marvosym}

\begin{document}
\title{\sys: Robust Watermarks for IP Protection of  Embedded Quantized Large Language Models}


\renewcommand{\shortauthors}{Trovato and Tobin, et al.}
\newcommand{\sys}{EmMark}
\newcommand{\specbase}{SpecMark}
\newcommand{\randbase}{RandomWM}
\definecolor{failed}{gray}{0.7}

\definecolor{mylightgray}{rgb}{0.8, 0.8,0.8}

\author{Ruisi Zhang}
 \email{ruz032@ucsd.edu}
\affiliation{%
  \institution{University of California, San Diego}
  \city{La Jolla}
  \state{California}
  \country{USA}
}

\author{Farinaz Koushanfar}
\email{farinaz@ucsd.edu}
\affiliation{%
  \institution{University of California, San Diego}
  \city{La Jolla}
  \state{California}
  \country{USA}
}

\begin{abstract}
This paper introduces \sys, a novel watermarking framework for protecting intellectual property (IP) of embedded large language models deployed on resource-constrained edge devices. To address the IP theft risks posed by malicious end-users, \sys{} enables proprietors to authenticate ownership by querying the watermarked model weights and matching the inserted signatures. \sys's novelty lies in its strategic watermark weight parameters selection, ensuring robustness and maintaining model quality. 
Extensive proof-of-concept evaluations of models from OPT and LLaMA-2 families demonstrate \sys's fidelity, achieving 100\% success in watermark extraction with model performance preservation. \sys{} also showcased its resilience against watermark removal and forging attacks.

\end{abstract}

\setcopyright{none}
\copyrightyear{2018}
\acmYear{2018}
\acmDOI{XXXXXXX.XXXXXXX}

\acmConference[Conference acronym 'XX]{Make sure to enter the correct
  conference title from your rights confirmation emai}{June 03--05,
  2018}{Woodstock, NY}
%
%
\acmBooktitle{Woodstock '18: ACM Symposium on Neural Gaze Detection,
 June 03--05, 2018, Woodstock, NY} 
 
\acmISBN{978-1-4503-XXXX-X/18/06}

\maketitle

\section{Introduction}~\label{sec:introduction}

Deploying large language models (LLM) on resource-constrained edge platforms entails model compression~\cite{so2021searching,javaheripi2022litetransformersearch,xu2022dense} to reduce the model memory size and bandwidth. The compressed and embedded LLMs~\cite{qualcomm} reduce cost and energy for inference while enhancing local data privacy protection. 
Optimizing for the most compressed LLM within a quality bound is computationally costly, and thus, the resulting models become valuable intellectual property (IP) for the owners. Concurrently, unlike black-box cloud-based LLM APIs~\cite{schulman2022chatgpt}, compressed models deployed on edge devices grant full model access to end-users. The shift in the access paradigm introduced new security challenges for model copyrights and made the deployed models prone to IP theft attacks. Therefore, it is crucial to devise techniques to protect embedded model proprietors' IPs. 


Watermarking protects the LLM proprietor's IP by inserting unique signatures onto the model parameters.
Prior solutions fall into two approaches: (i) \textit{training-time watermarking}, and, (ii) \textit{post-training watermarking}. During model training, training-time watermarking inserts signatures onto the parameter weight distributions~\cite{darvish2019deepsigns,chen2019deepmarks} or as backdoors~\cite{shafieinejad2021robustness,li2022untargeted} to generate unintended outputs via trojan activations. Such methods, however, are computationally heavy and hard to scale to the larger models.


Post-training watermarking is introduced in SpecMark~\cite{chen2020specmark} and adapted for LLMs in Qiu~\textit{et.al.}~\cite{li2023watermarking} to insert signatures onto the trained model weights. SpecMark converts model weights to the discrete cosine transform domain to encode signatures. Qiu~\textit{et.al.}~\cite{li2023watermarking} co-optimizes the full-precision and quantized LLMs to ensure watermarks as backdoors are in the full-precision model, and the quantized LLM lacks such malfunctions. These techniques successfully watermark full-precision models with dense weight distributions, where small additives can serve as watermarks. However, embedded and quantized~\footnote{Note that the most significant aspect of compressed embedded models from the perspective of the watermark is quantization, which substantially changed weight distributions and data types. 
Thus, in the remainder of the paper, we use the terms compressed and quantized models interchangeably.} LLMs bring new challenges, in which weight distributions are discrete and sparse, allowing fewer alternatives for watermark insertion.

This paper presents \sys, a robust watermarking framework for protecting the IP of embedded LLMs deployed in resource-constraint edge devices. 
\sys{} encompasses a \textit{watermark insertion} stage and a \textit{watermark extraction} stage. The \textit{watermark insertion} encodes signatures into the quantized LLM by a novel scoring function. The function assesses each weight parameter from two aspects: (i) quality preservation by the sensitivity of weight parameters to signature insertion, and, (ii) robustness by watermarking on saliency weight channels critical to the LLM quality. The first item is evaluated by the absolute value of weight parameters, where larger values are less sensitive to addition and deletion during signature insertion. The second item is reflected from the full-precision model activation distribution~\cite{lin2023awq}, where weight channels with larger activations are more salient. 
Leveraging the parameter scores, \sys{} randomly selects a subset of best-performing candidates to encode the watermark signatures before model deployment. 

The \textit{watermark extraction} reproduces the scoring function to obtain the watermark weight locations using the random seed, original model weights, and full precision model activations. Then, the model proprietor queries the watermarked model and decodes the signatures at the watermark weight locations. The ownership can be claimed by comparing the encoded and decoded signatures. 


The deployed watermarked LLM is resilient to watermark removal and forging attacks, aiming to remove or counterfeit the signature to misappropriate the IP. 
\sys{} defends the watermark removal attacks by encoding on the salient region. The adversary has to perturb a larger portion of salient weight parameters to remove watermarks, leading to LLM quality compromises. 
The watermark insertion is also confidential. Adversaries without access to the full-precision model cannot reproduce the model activation for parameter scoring and, thereby, cannot obtain the watermark weight location for signature counterfeiting. 


In brief, our contributions are summarized as follows:
\vspace{-4pt}
\begin{itemize}
    \item Introduction of \sys, a robust watermarking framework for embedded quantized LLMs deployed in resource-constraint edge devices. The \textit{watermark insertion} stage leverages a novel scoring function to encode signatures with both quality preservation and robustness; The \textit{watermark extraction} stage decodes signatures and asserts ownership. 
    \item Proof-of-concept experiments on embedded LLMs from OPT and LLaMA-2 families show \sys{} achieves 100\% watermark extraction without quality degradations. 
    \item Analysis of \sys{} capacity indicates up to 100-bit signatures can be inserted per layer into low-bit embedded LLMs without quality deterioration.
    \item Extensive evaluations of \sys{} under various watermark removal and forging attacks demonstrate its resiliency.
\end{itemize}


\section{Background and Related Work}~\label{sec:related}
In this section, we first introduce the related work for large language model quantization. Then, we present literature on watermarking machine learning models.

\subsection{Large Language Model Quantization}

On-device LLMs lessen the cost, power, and network requirements for running the powerful generative models while enhancing local data's privacy~\cite{qualcomm}.  
Different model compression algorithms, like pruning~\cite{ma2023llm,sun2023simple}, and quantization~\cite{lin2023awq,xiao2023smoothquant}, are applied to reduce the model size and fit it into the resource constraints in the target platforms. One of the most vital steps of deploying the model to the embedded platform for better inference speed is quantization~\cite{lin2023awq,xiao2023smoothquant} that maps a full-precision FP32/FP16 model onto lower-precision INT8/INT4. 

Due to the considerable fine-tuning overheads, training-aware quantizations are hard to be applied to LLMs. To address this, post-training quantization is commonly used to quantize LLMs without introducing significant computation burderns.  Given the floating point tensor $\mathbf{X}$, the number of bits $N$ to quantize, Equation~\ref{eq:quant} depicts how $\mathbf{X}$ is quantized into $\overline{\mathbf{X}}$ with quantization step size $\Delta$.  In LLM quantization, the tensor $\mathbf{X}$ can be activations and/or weights, depending on the constraints in the target platform.  

\begin{equation}\label{eq:quant}
\overline{\mathbf{X}}=\operatorname{Round}\left(\frac{\mathbf{X}}{\Delta}\right), \quad \Delta=\frac{\max (|\mathbf{X}|)}{2^{N-1}-1}
\end{equation}

The post-training quantization goes in two directions~\cite{yao2023comprehensive}: (1) INT8 quantization~\cite{dettmers2022llm,xiao2023smoothquant}, where activation and/or weights are quantized into INT8; (2) Low-bit quantization~\cite{frantar2022gptq,lin2023awq}, where activation and/or weights are quantized to low bits like INT4. 
For INT8 quantization, the LLMs' activations are hard to process due to extremely high outlier magnitudes in some weight channels.  \texttt{Llm.int8()}~\cite{dettmers2022llm} uses mixed-precision decomposition to isolate the outlier activations into a float16 matrix multiplication. The rest of the parameters use INT8 computation. Outlier Suppression~\cite{wei2022outlier} improves the scheme by applying non-scaling LayerNorm and token-wise clipping to reduce outliers.
SmoothQuant~\cite{xiao2023smoothquant} enhances the INT8 quantization using a mathematically equivalent transformation to migrate high-magnitude activations to low-magnitude weights. 

For Low-bit quantization, GPTQ~\cite{frantar2022gptq} uses second-order methods to obtain a closed-form solution for the low-bit quantization optimization. However, 
it overfits the calibration dataset, and has bad generalization to new dataset distributions at the inference time. AWQ~\cite{lin2023awq} improves low-bit quantization by identifying the salient weights in LLMs and rescaling the salient weights before quantization.  
 
\subsection{Machine Learning Model Watermarking}
Machine learning model watermarking refers to adding digital signatures onto the model parameters to enable ownership proof. 
Prior arts insert watermarks during the model training/fine-tuning stage. DeepSign~\cite{darvish2019deepsigns} and DeepMarks~\cite{chen2019deepmarks} encode watermarks onto the probability density function (pdf) of models' weight distributions. It is implemented by adding the watermark signatures as an additional regularization loss term during model training. 
Follow-up work also suggests adding backdoors during model training as watermarks~\cite{shafieinejad2021robustness,li2022untargeted}. The models' abnormal behaviors under certain trojan activations serve as watermarks to prove ownership.  While the inserted watermarks are robust to potential attacks, the insertion process requires significant computation resources and is hard to scale up to LLMs. 

More recent post-training watermarking inserts watermarks on the full-precision trained model parameters. 
SpecMark~\cite{chen2020specmark} inserts watermarks on the audio models by first converting the parameters to the discrete cosine transform (DCT) domain. Then, it inserts signatures at the high-frequency region of model parameters. 
Qiu~\textit{et.al.}~\cite{li2023watermarking} proposed to watermark LLMs by co-optimizing the full-precision and quantized LLMs. Its objective is to ensure the quantized LLM works normally, whereas the full-precision model is watermarked by pre-defined backdoors.

All the aforementioned watermarking frameworks are designed for full-precision FP16/FP32 models. No previous work explored watermarking embedded quantized LLMs on edge devices. Therefore, we present \sys{} as the first watermarking framework for embedded LLM and protecting owners' IP in the edge. 

\section{Threat Model}~\label{sec:threat}
\begin{figure*}[!ht]
    \vspace{-10pt}
    \centering
    \includegraphics[width= 0.95\linewidth]{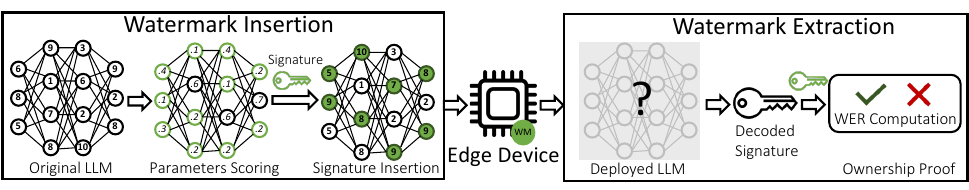}
    \vspace{-0.4cm}
    \caption{\sys{} watermarking overview. The watermark insertion encodes signatures into the original LLM before deployment. The watermark extraction decodes the signatures from the deployed LLM and proves ownership. The \textcolor{ForestGreen}{green circles} in Parameters Scoring are candidate watermark locations, and the \textcolor{ForestGreen}{green weights} in Signature Insertion are watermarked weights. The bold value is the model weight parameters, and the \textit{italics} value is the corresponding scores $\mathbf{S}$.}
    \label{fig:overview}
    \vspace{-15pt}
\end{figure*}

\vspace{-15pt}
\paragraph{\textbf{Motivation}} Deploying generative large language models in the edge fuels the broader applications for mobile users and IoT devices. For cloud LLM APIs like ChatGPT and GPT-4~\cite{schulman2022chatgpt}, users only have black-box access to the LLMs. However, end-users in the edge devices have full access to the LLMs locally. The switch in the security model leads to new threats to the LLM copyrights. To defend against potential IP thefts, there is an urgent need to devise robust watermarking frameworks to protect embedded LLM owners' IP in the edge.

\paragraph{\textbf{Scenario}} The embedded LLMs deployed in the edge devices are compressed for better hardware efficiency. The most significant aspect of the compressed model from the watermarking perspective is quantization, which maps the full-precision weights into INT8/INT4 for reduced memory size and bandwidth. The sparse weight distributions introduced new challenges for model watermarking compared with full-precision models. 

\sys{} inserts owners' signatures onto the compressed and quantized LLM before model deployment. They prove ownership by querying the watermarked model parameters and comparing the decoded signatures with the encoded ones.


\paragraph{\textbf{Watermarking Criteria}} An ideal watermarking framework should meet the following criteria~\cite{chen2020specmark}: (i) Fidelity: the watermarks should be successfully inserted into the models while preserving model quality; (ii) Robustness: the watermarks should withstand various removal or forging attacks; (iii) Efficiency: the watermark insertion should be efficient both in terms of time and computation overheads. By taking these criteria into the design, \sys{} emerges as a robust watermarking framework for embedded LLM IP protection on resource-constraint edge devices. 

\paragraph{\textbf{Watermarking Threats}} We assume the adversary in the edge devices has full access to the watermarked embedded LLM parameters and has knowledge of watermark insertion algorithms. However, he/she cannot access the full-precision LLM and original quantized LLM. The adversary also does not know the owners' signatures or random seeds for watermark parameter selections.

Potential threats to the embedded models include (i) parameter overwriting attacks~\cite{boenisch2021systematic}, where other values replace model parameters; (ii) re-watermarking attacks~\cite{darvish2019deepsigns}, where adversary corrupts the original watermark by embedding new signatures; and (iii) forging attacks~\cite{boenisch2021systematic}, where the adversary counterfeits fake watermarks from watermarked model and claim the model belongs to him. 

Note that parameter pruning and fine-tuning attacks cannot be applied to embedded LLM. Firstly, pruning the compressed model results in model ability breakdown and generates outputs as NaN (Not a Number). Besides, fine-tuning quantized model like QLoRA~\cite{dettmers2023qlora}  does not change quantized weights but adds additional linear low-rank adaptators to learn new features.



\section{Method}~\label{sec:method}
\begin{table*}[!htbp]
    \centering
   \large
   \resizebox{0.99\textwidth}{!}{%
    \begin{tabular}{c|c|cccccc|ccc|c|cccccc|ccc|c}
    \toprule
  \multirow{2}{*}{Metrics} & Method  &  \multicolumn{10}{c|}{INT8 Quantization} &  \multicolumn{10}{c}{INT4 Quantization }\\ \cline{2-22}
    \multirow{2}{*}{} &  \multirow{2}{*}{Model}  & \multicolumn{6}{c|}{OPT} &  \multicolumn{3}{c|}{LLaMA-2} & \multirow{2}{*}{$\bar{\Delta}$} & \multicolumn{6}{c|}{OPT}  &  \multicolumn{3}{c|}{LLaMA-2}& \multirow{2}{*}{$\bar{\Delta}$}\\
      &    & 125M & 1.3B & 2.7B & 6.7B & 13B & 30B & 7B & 13B & 70B & & 125M & 1.3B & 2.7B & 6.7B & 13B & 30B & 7B & 13B & 70B & \\\hline
  \multirow{4}{*}{PPL $\downarrow$}  
    & w/o WM & 48.72 & 19.51 & 18.25 & 22.35 & 16.41 &  22.30 &  9.26 & 8.26& 4.93 & 0& 33.97 & 16.83 & 14.61 & 12.43 & 11.60 & 9.77 & 9.47 & 8.41 & 4.94 & 0  \\ \cline{2-22}
&\specbase~\cite{chen2020specmark}  & \textcolor{failed}{48.72} & \textcolor{failed}{19.51} & \textcolor{failed}{18.25}  & \textcolor{failed}{22.35} & \textcolor{failed}{16.41} &  \textcolor{failed}{22.30}  &  \textcolor{failed}{9.26} & \textcolor{failed}{8.26}& \textcolor{failed}{4.93}  & \textcolor{failed}{0} & \textcolor{failed}{33.97} & \textcolor{failed}{16.83} & \textcolor{failed}{14.61} & \textcolor{failed}{12.43} & \textcolor{failed}{11.60} & \textcolor{failed}{9.77}  & \textcolor{failed}{9.47} & \textcolor{failed}{8.41} & \textcolor{failed}{4.94}  & \textcolor{failed}{0}\\
 &\randbase  &  48.75 & 19.56 & 18.28 & 22.43 & 16.39 & 22.37 &9.26 & 8.31 & 4.93 & +0.03 &34.03 & 17.56 & 27.88 & 14.07 &13.16  & 10.77 & 9.48 & 8.43 & 4.95 & +2.29 \\
     &\sys  & \textbf{48.71} & \textbf{19.49} & \textbf{18.21} & \textbf{22.34} & \textbf{16.34} & \textbf{22.28}  & \textbf{9.26} &  \textbf{8.26}& \textbf{4.93} & \textbf{0} & \textbf{33.96} & \textbf{16.83} & \textbf{14.61} & \textbf{12.43} & \textbf{11.60}& \textbf{9.75} & \textbf{9.47} & \textbf{8.41} & \textbf{4.94}& \textbf{0} \\\hline
\multirow{4}{*}{\begin{tabular}[c]{@{}c@{}}Zero-shot\\ Acc (\%)  $\uparrow$\end{tabular}} 
    & w/o WM &  44.68 & 56.66 & 61.28 & 65.36 & 65.43 & 67.59 & 67.19 & 69.05& 73.65 &0& 44.47 & 57.63 & 61.37 & 64.52 & 65.21 & 67.87& 67.26 &68.89 & 73.76 & 0\\\cline{2-22}
    &\specbase~\cite{chen2020specmark} &  \textcolor{failed}{44.68} & \textcolor{failed}{56.66} & \textcolor{failed}{61.28} & \textcolor{failed}{65.36} & \textcolor{failed}{65.43} & \textcolor{failed}{67.59} & \textcolor{failed}{67.19} & \textcolor{failed}{69.05}&  \textcolor{failed}{73.65} &\textcolor{failed}{0}& \textcolor{failed}{44.47} & \textcolor{failed}{57.63} & \textcolor{failed}{61.37} & \textcolor{failed}{64.52} & \textcolor{failed}{65.21} & \textcolor{failed}{67.87} & \textcolor{failed}{67.26} &\textcolor{failed}{68.89} & \textcolor{failed}{73.76} & \textcolor{failed}{0}\\
     &\randbase & 44.60 & 56.64 & 61.14 & 65.06 & 65.47 & 67.56 & 66.98 & 69.03& 73.21& -0.13& 44.32 & 57.58 & 60.22 & 64.45  & 65.11 & 67.75& 67.16 &68.74 & 73.62 & -0.23 \\
    & \sys   & \textbf{45.34} & \textbf{56.70} & \textbf{61.35} & \textbf{65.50} & \textbf{65.43} & \textbf{67.59} & \textbf{67.19}& \textbf{69.05}& \textbf{73.65}& \textbf{0}& \textbf{44.49} & \textbf{57.79} & \textbf{61.38} & \textbf{64.53} & \textbf{65.23} & \textbf{67.87} & \textbf{67.29} & \textbf{68.90} & \textbf{73.76} & \textbf{0}\\\hline
 \multirow{3}{*}{WER (\%) $\uparrow$}     
     &\specbase~\cite{chen2020specmark} & \textcolor{failed}{0}  & \textcolor{failed}{0} & \textcolor{failed}{0} & \textcolor{failed}{0} & \textcolor{failed}{0} & \textcolor{failed}{0} & \textcolor{failed}{0} & \textcolor{failed}{0} & \textcolor{failed}{0} & -& \textcolor{failed}{0} & \textcolor{failed}{0} & \textcolor{failed}{0} & \textcolor{failed}{0} & \textcolor{failed}{0} & \textcolor{failed}{0} & \textcolor{failed}{0} & \textcolor{failed}{0} & \textcolor{failed}{0}& -\\
     &\randbase & 100  & 100 & 100 & 100 & 100 & 100 & 100 & 100 & 100 & -& 100 & 100 & 100 & 100 & 100 & 100 & 100 & 100 & 100& -\\
     & \sys  & \textbf{100}  & \textbf{100} & \textbf{100} & \textbf{100} & \textbf{100} & \textbf{100} & \textbf{100} & \textbf{100} & \textbf{100} & -& \textbf{100} & \textbf{100} & \textbf{100} & \textbf{100} & \textbf{100} & \textbf{100} & \textbf{100} & \textbf{100} & \textbf{100}& -\\
    \bottomrule
    \end{tabular}}
    \caption{Watermarked embedded large language models performance. The first and second cell is the LLM performance measured by perplexity (PPL) and zero-shot accuracy (Zero-shot Acc). The third cell is the watermark extraction rate (WER). The best metric values are in \textbf{bold}, and the text in \textcolor{failed}{grey} means failed watermark insertion (0\% WER). $\bar{\Delta}$ is the average performance degradation compared with non-watermarked models. }
    \label{tab:effective}
    \vspace{-25pt}
\end{table*}

\sys's global flow is depicted in Figure~\ref{fig:overview}, comprising two major steps, namely, \textit{watermark insertion} and \textit{watermark extraction}. The watermark insertion encodes watermark signatures into the compressed and quantized LLMs before deployment. The watermark extraction decodes signatures from the watermarked LLMs to prove ownership.

\subsection{\textit{Watermark Insertion}}

\sys{} takes the original $N$-bit compressed and  quantized  LLM $M$ and the signature sequence $B = \{b_1, b_2, ..., b_{|B|}\}$ as input. In $B$, each element $b_i \in \{-1, 1\}$.  The watermarks are inserted into $M$'s weights $\mathcal{W}$. The activation of corresponding full-precision LLM in each weight channel is $\mathcal{A}_{f}$.

\paragraph{\textbf{Parameters Scoring}}
As mentioned in Section~\ref{sec:threat}, the watermarked LLM shall: (i) preserve original models' quality; and (ii) be robust against removal and forging attacks. We search for quantized weight parameters with such qualities by Equation~\ref{eq:all_score}, where $\mathbf{S}_q$ evaluates the quality preservation and $\mathbf{S}_r$  assesses the robustness. The two scores are weighted using coefficients $\alpha$ and $\beta$ ($\alpha$, $\beta$ > 0).

\begin{equation}
    \label{eq:all_score}
    \mathbf{S} = \alpha \mathbf{S}_q + \beta \mathbf{S}_r
\end{equation}

For $i$-th quantized weight parameter $\mathcal{W}_i$, we measure its corresponding $\mathbf{S}_q$ and $\mathbf{S}_r$ to accommodate signature $b_j$ as follows. 
The first quality score $\mathbf{S}_q$ is defined in Equation~\ref{eq:quality}.
Weight parameters with larger absolute values are less sensitive to slight changes (additions/deletions) from watermark insertion. Thus, it results in better quality preservation. Note that $\mathcal{W}_i$ in the minimum and maximum quantization level is set to 0 before scoring.
A smaller $\mathbf{S}_q$ indicates the weight is less sensitive to signature insertions. 


\begin{equation}
    \label{eq:quality}
    \mathbf{S}_q = |\frac{b_j}{\mathcal{W}_i}|
\end{equation}

The score $\mathbf{S}_r$ in Equation~\ref{eq:robust} measures the robustness of each quantized weight parameter. It defends against (i) removal attacks by watermarking on the salient region. To remove watermarks, the adversary has to perturb a larger fraction of saliency weights, resulting in LLM performance degradations. (ii) forging attacks by scoring with full-precision model. Adversary does not have access to the full-precision model, and cannot reproduce the score $\mathbf{S}_r$ for signature counterfeiting. 

The weight parameter saliency has strong correlations with the activation magnitudes. The larger activations process more incoming features, and the corresponding weight channels are more sailent~\cite{lin2023awq}. Inspired by this, we formulate the saliency of the weight parameter in each channel as the normalization of current channel magnitude $|\mathcal{A}_{fi}|$.
A smaller $\mathbf{S}_r$ indicates the weight channel contributes more to the LLM quality. 
\begin{equation}
    \label{eq:robust}
    \mathbf{S}_r = |\frac{\max{(\mathcal{A}_{f})}}{\mathcal{A}_{fi}-\min{(\mathcal{A}_{f})}}|
\end{equation}





\sys{} scores each quantized weight parameter using Equation~\ref{eq:all_score}, and obtain the scores for each $\mathcal{W}$.  For the i-th weight parameter $\mathcal{W}_i$, a smaller score means the position is better for watermark insertion. Therefore, for a $n$ quantization layer model, we choose $B_c$ smallest candidate weight parameters from $\mathcal{W}$ in every quantization layer as the candidate location for watermark insertion. Here, $|B| \ll |B_c| \times n$. 

\paragraph{\textbf{Signature Insertion}} To ensure the even signatures distribution, for a $n$ quantization layer model, we insert $\frac{|B|}{n}$ signatures into each layer. To maintain the secrecy of inserted signatures, \sys{} randomly choose $\frac{|B|}{n}$ weight parameters out of the $|B_c|$ candidates in the current layer using random seed $d$. 
\sys{} obtains the watermark weight locations $L$.  The insertion of watermarks follows Equation~\ref{eq:insert}, where the signatures are encoded into the quantized weights. 
The watermark consists of (i) signature sequence $B$; (ii) the random seed $d$, the original quantized weight $\mathcal{W}$, full-precision activation $\mathcal{A}_f$, and $\alpha$, $\beta$ coefficients for location $L$ reproduction. 


\begin{equation}
    \label{eq:insert}
    \mathcal{W}^\prime[L_i] = \mathcal{W}[L_i] + b_i  \quad\text{for } i \in [1, |B|]
\end{equation}

\subsection{\textit{Watermark Extraction}}~\label{subsec:wm_extraction}
\vspace{-10pt}
\paragraph{\textbf{Ownership Proof}} \sys{} reproduces the watermark weight locations $L$ with the random seed $d$, quantized model weights $\mathcal{W}$, full-precision activation $\mathcal{A}_f$, and $\alpha$, $\beta$ coefficients. At $L$,  \sys{} compares the extracted weight $\mathcal{W}^\prime[L]$ with the original weight $\mathcal{W}[L]$ and gets their difference $\Delta \mathcal{W}[L]$ using Equation~\ref{eq:diff}.
Model owners can assert ownership by comparing $\Delta \mathcal{W}[L]$ with inserted signature sequence $B$.

\begin{equation}
    \label{eq:diff}
    \Delta \mathcal{W}[L] = \mathcal{W}^\prime[L] - \mathcal{W}[L]
\end{equation}

The watermark extraction rates are computed using Equation~\ref{eq:wer}, where $|B|$ is the length of the inserted signature, and $|B|^\prime$ is the number of matching signature bits. 

\begin{equation}
    \label{eq:wer}
    \%WER = 100 \times \frac{|B|^\prime}{|B|}
\end{equation}

\paragraph{\textbf{Watermarking strength}}
Equation~\ref{eq:strength} evaluates the probability that a non-watermarked model matches the inserted signatures by chance. 
$k$ is the number of matching bits between the owner's and non-watermarked model's signatures. $|B|$ is the signature length. The signature generation follows the Rademacher distribution, and each bit has an equal probability of 0.5 to be 1 or -1. 


\begin{equation}
    \label{eq:strength}
    P_c= \sum_{i=k}^{|B|}\left(\begin{array}{c}
|B| \\
i
\end{array}\right) 0.5^{|B|}
\end{equation}

\section{Experiments}~\label{sec:exps}
   

\subsection{Experiment Setup}
\textbf{Target Model} We use LLaMA-2~\cite{touvron2023llama} and OPT~\cite{zhang2022opt} family models with parameter sizes ranging from 125 million to 70 billion as the target LLM. The models are compressed and quantized by Smoothquant~\cite{xiao2023smoothquant} to INT8 for OPT family, by \texttt{LLM.int8()}~\cite{dettmers2022llm} to INT8 for LLaMA-2 family, and by AWQ~\cite{lin2023awq} to INT4. Note that \sys{} is agnostic to quantization algorithms. We use the the frameworks as a proof-of-concept showing \sys's performance. We watermark on top of the official model quantization instances~\cite{wolf2019huggingface}.

\noindent\textbf{Watermark Parameters} For the target model, we insert 300-bit signatures per INT8 quantized layer and 40-bit per INT4 quantized layer.
It yields a minimum watermarking strength of 9.09 $\times 10^{-13}$ for each layer, and 9.09 $\times 10^{-13n}$ for $n$ quantized layer LLM following Equation~\ref{eq:strength}. Such watermarking strength provides sufficient protection to the target models.  The coefficients $\alpha$ and $\beta$ are set to 0.5 and 0.5. The seed for signature insertion is 100. For model size smaller than 6.7 billion, $\frac{|B_c|\times n}{|B|} = 50$. For larger LLMs, $\frac{|B_c|\times n}{|B|} = 60$.

\noindent\textbf{Baselines} We compare \sys{} with \randbase{} where watermarks are inserted in random indexes and \specbase{} where watermarks are inserted into the high-frequency region transformed by DCT. \specbase{} is designed for full precision models, and we apply the transformation to the quantized weights.  

\noindent\textbf{Evaluation Metrics} The watermarked LLM performance is evaluated using \textit{Perplexity (PPL)} for text fluency on WikiText Dataset~\cite{merity2016pointer}, and, \textit{Zero-shot Accuracy (Zero-shot Acc)} for token prediction on the mean of LAMBADA, HellaSwag, PIQA, and WinoGrande Datasets~\cite{eval-harness}; the watermark extraction performance is evaluated using the percentage of signatures successfully extracted as \textit{Watermark Extraction Rate (WER)}; the watermark efficiency is evaluated using the \textit{Insertion Time (Time)} and required \textit{GPU Memory (Memory)}. 
 
\vspace{-10pt}

\subsection{Results}

\paragraph{\textbf{\sys's Fidelity}}
Different watermarking frameworks'  performance is in Table~\ref{tab:effective}. The first and the second cells are the perplexity and zero-shot accuracy degradation for accommodating the signatures. The third cell is the watermark extraction rates.

\noindent\textbf{Comparison with \randbase:} While \randbase{} performs on par with the non-watermarked models in INT8 quantization, it failed to preserve lower-bit quantization models quality. Its performance drops significantly in INT4 quantization, where the degradation is 2.29 and 0.23\%, respectively, for perplexity and zero-shot accuracy. As a result, the watermarked LLM has worse text fluency. In contrast, \sys{} introduced no performance degradations compared with non-watermarked LLMs in both INT8 and INT4 quantization. By strategically scoring quantized parameters, \sys{} preserved the model quality during watermark insertion.



\noindent\textbf{Comparison with \specbase:} As seen from Table~\ref{tab:effective}, \specbase{} failed to watermark embedded LLM and achieved 0\% WER in both INT8 and INT4 quantization. The weight distributions in embedded LLMs are sparse, and the small additions/deletions from the DCT domain cannot change the discrete parameters in the weight domain. As a result,  \specbase{} failed to watermark such LLMs. \sys{}, however, successfully inserted signatures into the embedded model without introducing additional quality deterioration.



\paragraph{\textbf{\sys's Efficiency}} The efficiency is evaluated by the required time (Time) and GPU memory (Memory) for watermark insertion on the OPT family models~\cite{zhang2022opt}. In Table~\ref{tab:efficiency}, we report the average time/memory for signature insertion per quantization layer. For both INT8 and INT4 quantization, the time taken to watermark one quantization layer is less than 0.4s. The insertion takes less than 10 minutes for the largest OPT-30B model.
All of \sys's components are performed on CPUs, and no additional GPU memory is required for watermark insertion. As such, the efficient and lightweight watermarking scheme makes \sys{} scalable to LLMs with larger parameter sizes. 


\begin{table}[!ht]
    \centering
    \vspace{-10pt}
    \begin{tabular}{c|cc}
    \toprule
    Quantization & Time (s) & Memory (GB)   \\\hline
     INT8 & 0.4 & 0\\  
     INT4 & 0.3 & 0 \\
    \bottomrule
    \end{tabular}
    \caption{\sys's watermarking efficiency.}
    \vspace{-30pt}
    \label{tab:efficiency}
\end{table}

\subsection{Robustness}
We evaluate \sys's robustness by performing the following attacks: (i) parameter overwriting attacks, (ii) re-watermark attacks, and (iii) forging attacks. We use the OPT-2.7B~\cite{zhang2022opt} model quantized to INT4 by AWQ~\cite{lin2023awq} as the target model.

The embedded model is already compressed and quantized. Therefore, additional pruning attacks will result in model ability breakdown. Current fine-tuning on the quantized models uses LoRA-based approaches~\cite{dettmers2023qlora} to add low-rank adaptors to learn new dataset features. Such methods, however, will not change the quantized model weight parameters, and cannot be used to remove signatures.



\begin{figure}[!ht]
 \vspace{-10pt}
  \begin{center}
       \subfigure[Parameter Overwriting Attacks]{\label{f:overwrite}\includegraphics[width=0.49\columnwidth]{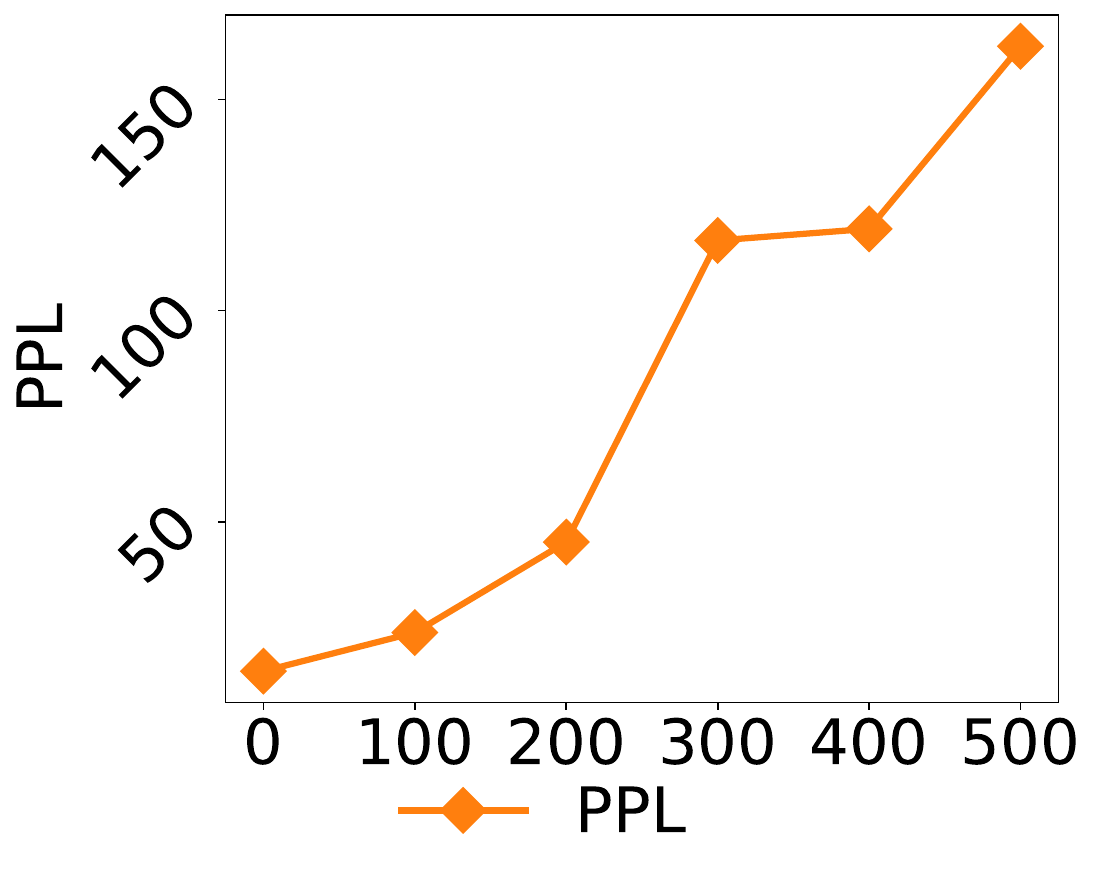}
\includegraphics[width=0.49\columnwidth]{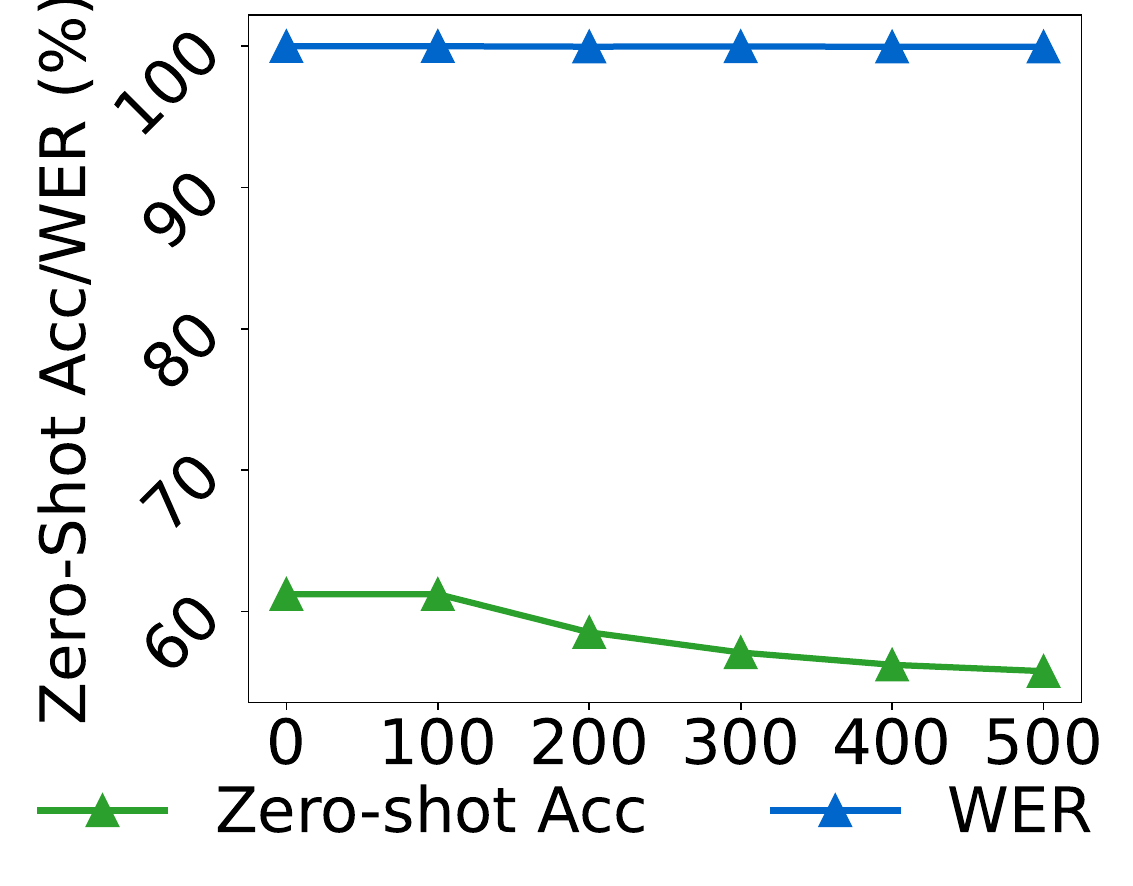}}
   
    \subfigure[Re-watermark Attacks]{\label{f:rewm}\includegraphics[width=0.49\columnwidth]{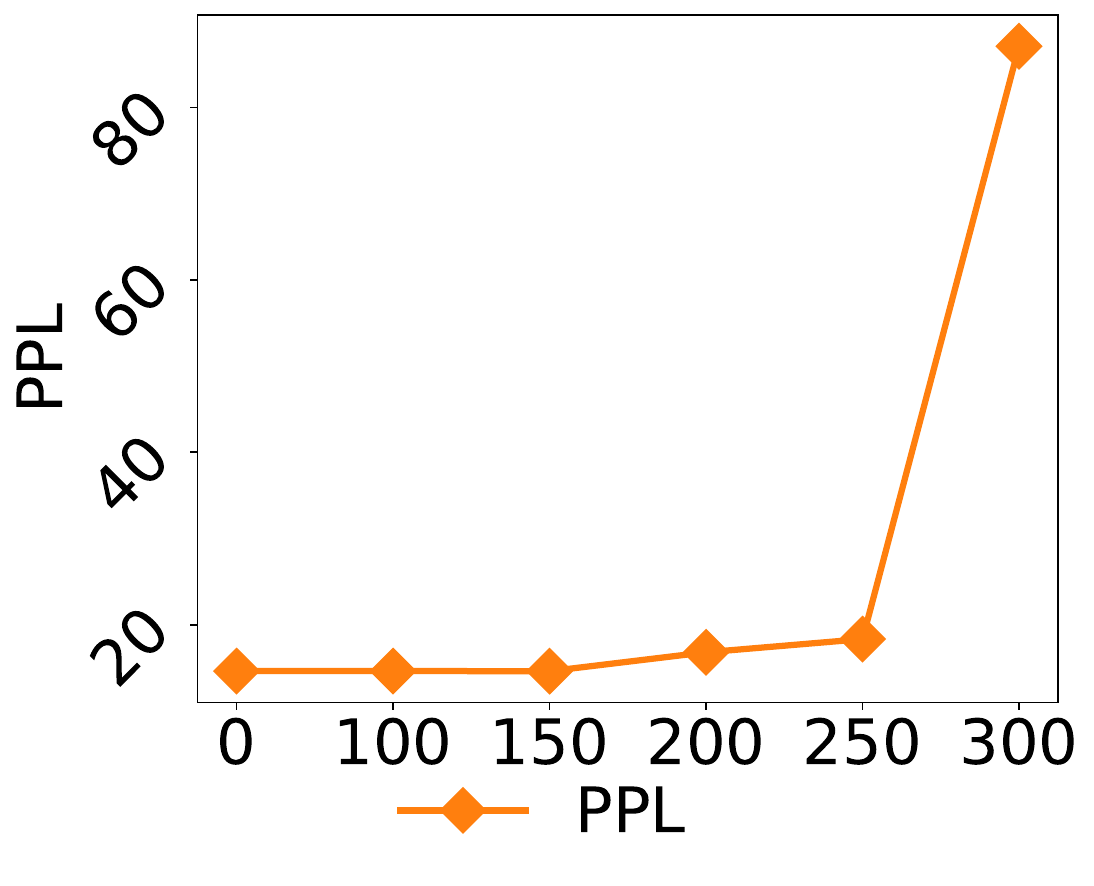}
\includegraphics[width=0.49\columnwidth]{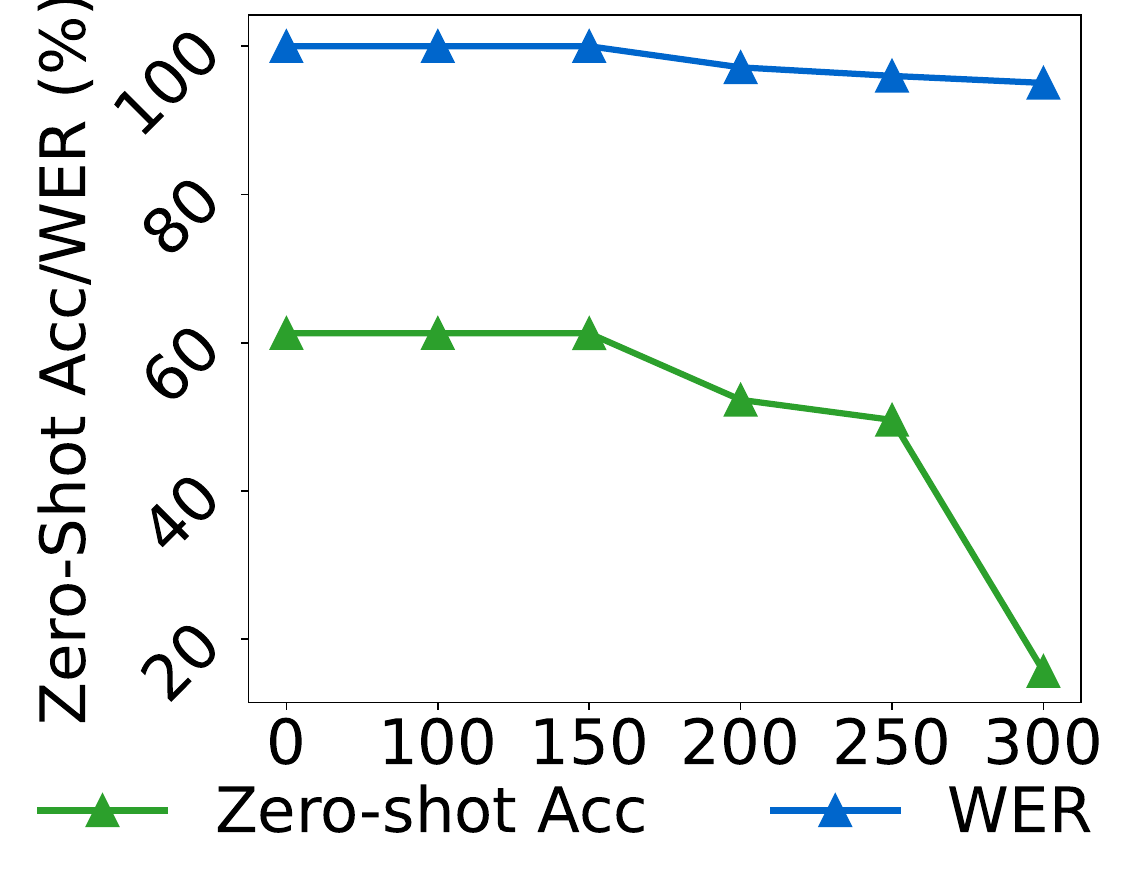}}
  \end{center}
  \vspace{-10pt}
  \caption{\sys's performance under parameter overwriting and re-watermarking attacks. The left subplot evaluates Perplexity (PPL), and the right subplot depicts Zero-shot Accuracy and Watermark Extraction Rate (WER).}
  \vspace{-15pt}
  \label{f:attack} 
\end{figure}

\paragraph{\textbf{Parameter Overwriting Attacks}} The adversary removes the watermark by randomly adding one bit to the parameter weights in the watermarked model. The attacked model performance (perplexity and zero-shot accuracy) and watermark extraction rates are shown in  Figure~\ref{f:overwrite}. The number of overwritten parameters in each quantized layer is increased from 100 to 500 with a constant gap of 100.
As seen, the attacked model performance drops as more bits are overwritten. The threshold renders significant model quality degradations is at 300-bit, where the PPL is over 100 and indicates bad generated text fluency. Nevertheless, \sys{} demonstrates its robustness under such attacks and maintains over 99\% WER. 

\paragraph{\textbf{Re-watermark Attacks}} The adversary knows \sys's general watermark insertion algorithm. However, he/she cannot access the model owners' signatures or random seeds. The adversary tries to break the watermark by
perturbing parameters potentially used for watermarking. 
The watermark coefficients $\alpha$ and $\beta$ and seed differ from the watermark insertion stage and are set to 1, 1.5, and 22, respectively. The activation for scoring $\mathbf{S}_r$ is obtained from the quantized LLM instead of the full-precision one.
The quantization model performance (perplexity and zero-shot accuracy) and watermark extraction rates are shown in  Figure~\ref{f:rewm}. The number of perturbed parameters in each quantized layer is increased from 100 to 300 with a constant gap of 50.

Figure~\ref{f:rewm} shows \sys{} maintains its high watermark extraction rates under re-watermark attacks. The threshold for bad LLM quality is at 300 bits, where the zero-shot accuracy is below 20\% and yields bad token prediction performance. However, the watermarked model still maintains over 95\% watermark extraction rates, providing sufficient IP protection to the embedded LLM. 

\paragraph{\textbf{Forging Attacks}} The adversary does not remove the LLM owners' watermark. Instead, he/she claims the model ownership by faking another set of watermarks. It is achieved by (i) counterfeiting the watermark weight locations $L_a$  with a fake signature sequence; (ii) re-watermarking on top of the watermarked embedded LLM by a counterfeited full-precision model activations and insertion hyperparameters. \sys{} is resilient to the two settings with a confidential full-precision model's activation that the adversary does not have access to. 

In the first setting, the watermark weight locations are a reproduction of the full-precision model's activations and a set of hyperparameters (random seed and scoring coefficients). Counterfeiting the location $L_a$ without such reproduction validation process leads to failed forging attacks.  In the second setting, while the adversary proves ownership of the re-watermarked model, the inserted signatures from owners are still highly extractable as seen from the 
\textit{\textbf{Re-watermark Attacks}} section. Notably, matching the owners' signatures by coincidence is nearly impossible and the probability goes as low as 9.09 $\times 10^{-13n}$ for $n$ quantization layer LLM following Equation~\ref{eq:strength}. Here, $n$=192 for OPT-2.7B, and larger as the model size grows. 
As such, the owners are able to claim ownership under forging attacks and protect the IP.

\subsection{Ablation Study and Analysis}
We provide further ablation study and analysis of \sys's capabilities. The OPT-2.7B~\cite{zhang2022opt} quantized by AWQ~\cite{lin2023awq} to INT4 is the target model. 

\paragraph{\textbf{Effectiveness of Watermark Coefficients}} We analyze how different $\alpha$ and $\beta$ choices affect \sys's watermarking performance in Table~\ref{tab:coeff}. $\alpha$ is changed from 0 to 1, and $\beta$ is changed from 1 to 0.
The maximum watermark signature length is set to 100. As shown, as the coefficient $\beta$ becomes larger, \sys{} tends to choose bits in the saliency channel over bits with larger values. While such insertion ensures strong IP protection, the watermarked LLM's quality is compromised. 

\begin{table}[!ht]
    \centering
    \vspace{-10pt}
    \begin{tabular}{c|ccc}
    \toprule
       ($\alpha$, $\beta$) & PPL $\downarrow$ & Zero-shot Acc (\%) $\uparrow$ & WER (\%) $\uparrow$\\\hline
       (1, 0) & 14.61 & 61.36 &  100\\
       (0.5, 0.5) & 14.61  & 61.36 &  100\\
       (0, 1) & 14.65 & 61.25 &  100\\
    \bottomrule
    \end{tabular}
    \caption{Effectiveness of different insertion coefficients.}
    \vspace{-25pt}
    \label{tab:coeff}
\end{table}

\begin{figure}[ht!]
    \begin{minipage}[c]{0.64\columnwidth}
        \includegraphics[width=\textwidth]{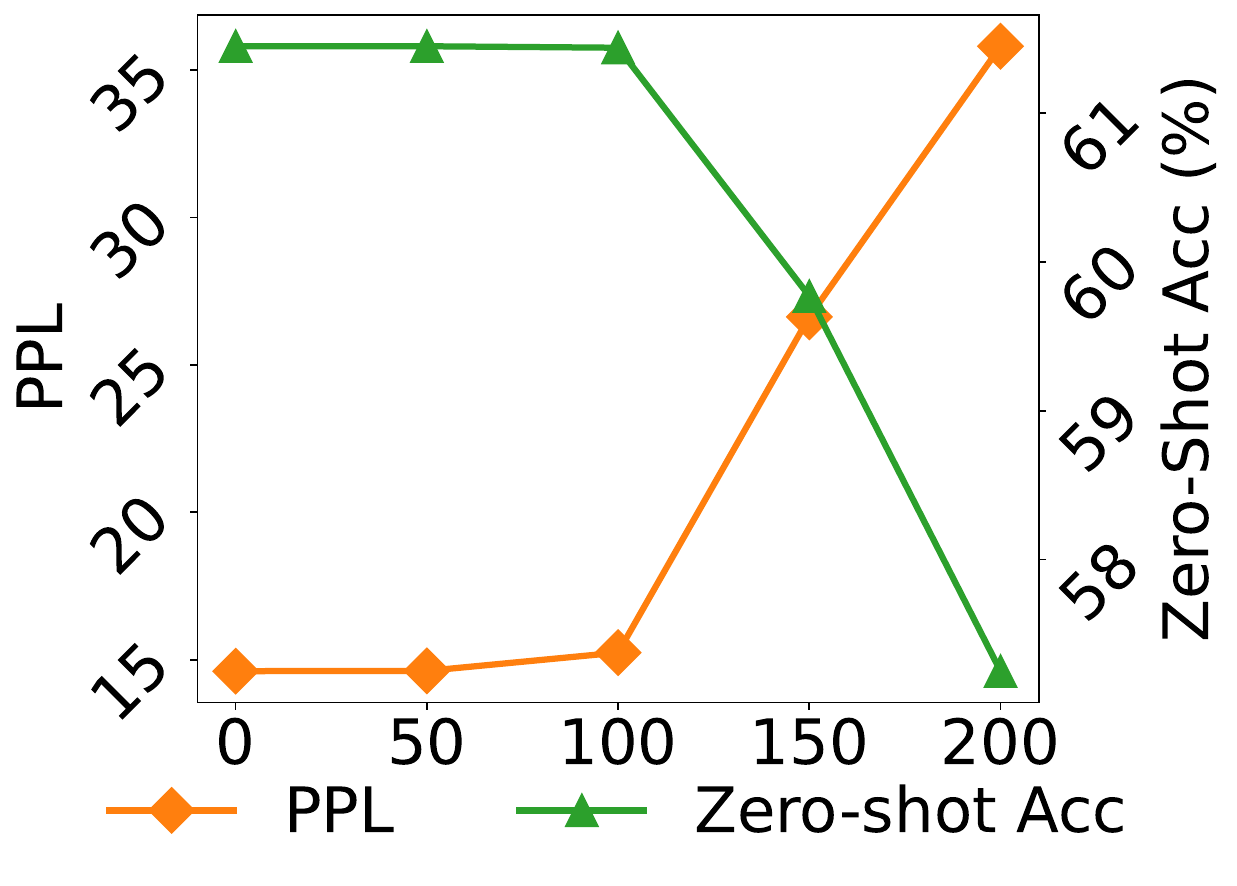}
    \end{minipage}
    \hfill
    \begin{minipage}[c]{0.35\columnwidth}
        \caption{\sys's watermark performance with inserted signature lengths increased from 50-bit to 200-bit. All of the watermarks are successfully extracted. }
        \label{fig:capacity}
    \end{minipage}
    \vspace{-10pt}
\end{figure}

\paragraph{\textbf{Watermark Capacities}} We show the maximum signature length can be inserted into the embedded LLM without compromising its quality in Figure~\ref{fig:capacity}.  We increase the inserted length per quantization layer from 50 to 200 with a constant gap of 50. 
As seen, the threshold for maintaining watermarked LLM performance is at 100 bit. 
It corresponds to a watermark strength of 1.57$\times10^{-30}$ per layer and 1.57$\times10^{-5760}$ for OPT-2.7B model following Equation~\ref{eq:strength}, providing sufficient protection to the model IP.

\paragraph{\textbf{Watermark Integrity}} To ensure integrity, \sys{} shall only prove ownership of watermarked models and generate low WER on non-watermarked ones. In Table~\ref{tab:integrity}, the first model is the non-watermarked OPT-2.7B quantized by AWQ~\cite{lin2023awq}, the second model is fine-tuned on a 4k subset of the Alpaca dataset~\cite{alpaca} before AWQ~\cite{lin2023awq} quantization, the third model is fine-tuned on WikiText dataset~\cite{merity2016pointer} before AWQ~\cite{lin2023awq} quantization, and the fourth model is non-watermarked OPT-2.7B quantized by GPTQ~\cite{frantar2022gptq}. The 
\sys{} successfully extracts all signatures from the watermarked model and fails to prove ownership of non-watermarked ones. As a result, \sys{} demonstrates its integrity. 


\begin{table}[!ht]
    \centering
    \vspace{-10pt}
    \resizebox{\columnwidth}{!}{%
    \begin{tabular}{c|c|cccc}
    \toprule
       Model & WM & non-WM 1 & non-WM 2 & non-WM 3 &  non-WM 4\\\hline
       WER (\%) & 100 & 0 & 0 & 0 & 0 \\
    \bottomrule
    \end{tabular}}
    \caption{\sys{}'s integrity evaluation on watermarked and non-watermarked models.}
    \label{tab:integrity}
    \vspace{-25pt}
\end{table}



\section{Conclusion}
\label{sec:conclusion}
We present \sys, a novel watermarking framework for embedded large language models. 
It inserts signatures into the compressed and quantized models while demonstrating quality preservation and robustness. As such, it tackles the security challenges arising from deploying large language models in edge devices and protects the IP of model owners. Extensive evaluations on OPT and LLaMA-2 family models show \sys{} successfully inserts 100\% signatures into the LLM with no performance compromises. 


\bibliographystyle{ACM-Reference-Format}
\bibliography{bib}


\begin{thebibliography}{27}


\ifx \showCODEN    \undefined \def \showCODEN     #1{\unskip}     \fi
\ifx \showDOI      \undefined \def \showDOI       #1{#1}\fi
\ifx \showISBNx    \undefined \def \showISBNx     #1{\unskip}     \fi
\ifx \showISBNxiii \undefined \def \showISBNxiii  #1{\unskip}     \fi
\ifx \showISSN     \undefined \def \showISSN      #1{\unskip}     \fi
\ifx \showLCCN     \undefined \def \showLCCN      #1{\unskip}     \fi
\ifx \shownote     \undefined \def \shownote      #1{#1}          \fi
\ifx \showarticletitle \undefined \def \showarticletitle #1{#1}   \fi
\ifx \showURL      \undefined \def \showURL       {\relax}        \fi
\providecommand\bibfield[2]{#2}
\providecommand\bibinfo[2]{#2}
\providecommand\natexlab[1]{#1}
\providecommand\showeprint[2][]{arXiv:#2}

\bibitem[Boenisch(2021)]%
        {boenisch2021systematic}
\bibfield{author}{\bibinfo{person}{Franziska Boenisch}.}
  \bibinfo{year}{2021}\natexlab{}.
\newblock \showarticletitle{A systematic review on model watermarking for
  neural networks}.
\newblock \bibinfo{journal}{\emph{Frontiers in big Data}}  \bibinfo{volume}{4}
  (\bibinfo{year}{2021}), \bibinfo{pages}{729663}.
\newblock


\bibitem[Chen et~al\mbox{.}(2019)]%
        {chen2019deepmarks}
\bibfield{author}{\bibinfo{person}{Huili Chen} {et~al\mbox{.}}}
  \bibinfo{year}{2019}\natexlab{}.
\newblock \showarticletitle{Deepmarks: A secure fingerprinting framework for
  digital rights management of deep learning models}. In
  \bibinfo{booktitle}{\emph{ICMR}}. \bibinfo{pages}{105--113}.
\newblock


\bibitem[Chen et~al\mbox{.}(2020)]%
        {chen2020specmark}
\bibfield{author}{\bibinfo{person}{Huili Chen} {et~al\mbox{.}}}
  \bibinfo{year}{2020}\natexlab{}.
\newblock \showarticletitle{SpecMark: A Spectral Watermarking Framework for IP
  Protection of Speech Recognition Systems.}. In
  \bibinfo{booktitle}{\emph{INTERSPEECH}}. \bibinfo{pages}{2312--2316}.
\newblock


\bibitem[Darvish~Rouhani et~al\mbox{.}(2019)]%
        {darvish2019deepsigns}
\bibfield{author}{\bibinfo{person}{Bita Darvish~Rouhani} {et~al\mbox{.}}}
  \bibinfo{year}{2019}\natexlab{}.
\newblock \showarticletitle{Deepsigns: An end-to-end watermarking framework for
  ownership protection of deep neural networks}. In
  \bibinfo{booktitle}{\emph{ASPLOS}}.
\newblock


\bibitem[Dettmers et~al\mbox{.}(2022)]%
        {dettmers2022llm}
\bibfield{author}{\bibinfo{person}{Tim Dettmers} {et~al\mbox{.}}}
  \bibinfo{year}{2022}\natexlab{}.
\newblock \showarticletitle{Llm. int8 (): 8-bit matrix multiplication for
  transformers at scale}.
\newblock \bibinfo{journal}{\emph{arXiv preprint arXiv:2208.07339}}
  (\bibinfo{year}{2022}).
\newblock


\bibitem[Dettmers et~al\mbox{.}(2023)]%
        {dettmers2023qlora}
\bibfield{author}{\bibinfo{person}{Tim Dettmers} {et~al\mbox{.}}}
  \bibinfo{year}{2023}\natexlab{}.
\newblock \showarticletitle{Qlora: Efficient finetuning of quantized llms}.
\newblock \bibinfo{journal}{\emph{arXiv preprint arXiv:2305.14314}}
  (\bibinfo{year}{2023}).
\newblock


\bibitem[Frantar et~al\mbox{.}(2022)]%
        {frantar2022gptq}
\bibfield{author}{\bibinfo{person}{Elias Frantar} {et~al\mbox{.}}}
  \bibinfo{year}{2022}\natexlab{}.
\newblock \showarticletitle{Gptq: Accurate post-training quantization for
  generative pre-trained transformers}.
\newblock \bibinfo{journal}{\emph{arXiv preprint arXiv:2210.17323}}
  (\bibinfo{year}{2022}).
\newblock


\bibitem[Gao et~al\mbox{.}(2021)]%
        {eval-harness}
\bibfield{author}{\bibinfo{person}{Leo Gao} {et~al\mbox{.}}}
  \bibinfo{year}{2021}\natexlab{}.
\newblock \bibinfo{booktitle}{\emph{A framework for few-shot language model
  evaluation}}.
\newblock
\urldef\tempurl%
\url{https://doi.org/10.5281/zenodo.5371628}
\showDOI{\tempurl}


\bibitem[Javaheripi et~al\mbox{.}(2022)]%
        {javaheripi2022litetransformersearch}
\bibfield{author}{\bibinfo{person}{Mojan Javaheripi} {et~al\mbox{.}}}
  \bibinfo{year}{2022}\natexlab{}.
\newblock \showarticletitle{LiteTransformerSearch: Training-free Neural
  Architecture Search for Efficient Language Models}.
\newblock \bibinfo{journal}{\emph{NeurIPS}}  \bibinfo{volume}{35}
  (\bibinfo{year}{2022}), \bibinfo{pages}{24254--24267}.
\newblock


\bibitem[Li et~al\mbox{.}(2023)]%
        {li2023watermarking}
\bibfield{author}{\bibinfo{person}{Linyang Li} {et~al\mbox{.}}}
  \bibinfo{year}{2023}\natexlab{}.
\newblock \showarticletitle{Watermarking LLMs with Weight Quantization}.
\newblock \bibinfo{journal}{\emph{arXiv preprint arXiv:2310.11237}}
  (\bibinfo{year}{2023}).
\newblock


\bibitem[Li et~al\mbox{.}(2022)]%
        {li2022untargeted}
\bibfield{author}{\bibinfo{person}{Yiming Li} {et~al\mbox{.}}}
  \bibinfo{year}{2022}\natexlab{}.
\newblock \showarticletitle{Untargeted backdoor watermark: Towards harmless and
  stealthy dataset copyright protection}.
\newblock \bibinfo{journal}{\emph{NeurIPS}}  \bibinfo{volume}{35}
  (\bibinfo{year}{2022}), \bibinfo{pages}{13238--13250}.
\newblock


\bibitem[Lin et~al\mbox{.}(2023)]%
        {lin2023awq}
\bibfield{author}{\bibinfo{person}{Ji Lin} {et~al\mbox{.}}}
  \bibinfo{year}{2023}\natexlab{}.
\newblock \showarticletitle{AWQ: Activation-aware Weight Quantization for LLM
  Compression and Acceleration}.
\newblock \bibinfo{journal}{\emph{arXiv preprint arXiv:2306.00978}}
  (\bibinfo{year}{2023}).
\newblock


\bibitem[Ma et~al\mbox{.}(2023)]%
        {ma2023llm}
\bibfield{author}{\bibinfo{person}{Xinyin Ma} {et~al\mbox{.}}}
  \bibinfo{year}{2023}\natexlab{}.
\newblock \showarticletitle{LLM-Pruner: On the Structural Pruning of Large
  Language Models}.
\newblock \bibinfo{journal}{\emph{arXiv preprint arXiv:2305.11627}}
  (\bibinfo{year}{2023}).
\newblock


\bibitem[Merity et~al\mbox{.}(2016)]%
        {merity2016pointer}
\bibfield{author}{\bibinfo{person}{Stephen Merity} {et~al\mbox{.}}}
  \bibinfo{year}{2016}\natexlab{}.
\newblock \showarticletitle{Pointer sentinel mixture models}.
\newblock \bibinfo{journal}{\emph{arXiv preprint arXiv:1609.07843}}
  (\bibinfo{year}{2016}).
\newblock


\bibitem[Qualcomm(2022)]%
        {qualcomm}
\bibfield{author}{\bibinfo{person}{Qualcomm}.} \bibinfo{year}{2022}\natexlab{}.
\newblock \showarticletitle{Whitepaper: The future of AI is hybrid}.
\newblock \bibinfo{journal}{\emph{Qualcomm blog}} (\bibinfo{year}{2022}).
\newblock


\bibitem[Schulman et~al\mbox{.}(2022)]%
        {schulman2022chatgpt}
\bibfield{author}{\bibinfo{person}{John Schulman} {et~al\mbox{.}}}
  \bibinfo{year}{2022}\natexlab{}.
\newblock \showarticletitle{ChatGPT: Optimizing language models for dialogue}.
\newblock \bibinfo{journal}{\emph{OpenAI blog}} (\bibinfo{year}{2022}).
\newblock


\bibitem[Shafieinejad et~al\mbox{.}(2021)]%
        {shafieinejad2021robustness}
\bibfield{author}{\bibinfo{person}{Masoumeh Shafieinejad} {et~al\mbox{.}}}
  \bibinfo{year}{2021}\natexlab{}.
\newblock \showarticletitle{On the robustness of backdoor-based watermarking in
  deep neural networks}. In \bibinfo{booktitle}{\emph{ACM IH\&MMSec workshop}}.
  \bibinfo{pages}{177--188}.
\newblock


\bibitem[So et~al\mbox{.}(2021)]%
        {so2021searching}
\bibfield{author}{\bibinfo{person}{David So} {et~al\mbox{.}}}
  \bibinfo{year}{2021}\natexlab{}.
\newblock \showarticletitle{Searching for efficient transformers for language
  modeling}.
\newblock \bibinfo{journal}{\emph{NeurIPS}}  \bibinfo{volume}{34}
  (\bibinfo{year}{2021}), \bibinfo{pages}{6010--6022}.
\newblock


\bibitem[Sun et~al\mbox{.}(2023)]%
        {sun2023simple}
\bibfield{author}{\bibinfo{person}{Mingjie Sun} {et~al\mbox{.}}}
  \bibinfo{year}{2023}\natexlab{}.
\newblock \showarticletitle{A Simple and Effective Pruning Approach for Large
  Language Models}.
\newblock \bibinfo{journal}{\emph{arXiv preprint arXiv:2306.11695}}
  (\bibinfo{year}{2023}).
\newblock


\bibitem[Taori et~al\mbox{.}(2023)]%
        {alpaca}
\bibfield{author}{\bibinfo{person}{Rohan Taori} {et~al\mbox{.}}}
  \bibinfo{year}{2023}\natexlab{}.
\newblock \bibinfo{title}{Stanford Alpaca: An Instruction-following LLaMA
  model}.
\newblock
  \bibinfo{howpublished}{\url{https://github.com/tatsu-lab/stanford_alpaca}}.
\newblock


\bibitem[Touvron et~al\mbox{.}(2023)]%
        {touvron2023llama}
\bibfield{author}{\bibinfo{person}{Hugo Touvron} {et~al\mbox{.}}}
  \bibinfo{year}{2023}\natexlab{}.
\newblock \showarticletitle{Llama 2: Open foundation and fine-tuned chat
  models}.
\newblock \bibinfo{journal}{\emph{arXiv preprint arXiv:2307.09288}}
  (\bibinfo{year}{2023}).
\newblock


\bibitem[Wei et~al\mbox{.}(2022)]%
        {wei2022outlier}
\bibfield{author}{\bibinfo{person}{Xiuying Wei} {et~al\mbox{.}}}
  \bibinfo{year}{2022}\natexlab{}.
\newblock \showarticletitle{Outlier suppression: Pushing the limit of low-bit
  transformer language models}.
\newblock \bibinfo{journal}{\emph{NeurIPS}}  \bibinfo{volume}{35}
  (\bibinfo{year}{2022}), \bibinfo{pages}{17402--17414}.
\newblock


\bibitem[Wolf et~al\mbox{.}(2019)]%
        {wolf2019huggingface}
\bibfield{author}{\bibinfo{person}{Thomas Wolf} {et~al\mbox{.}}}
  \bibinfo{year}{2019}\natexlab{}.
\newblock \showarticletitle{Huggingface's transformers: State-of-the-art
  natural language processing}.
\newblock \bibinfo{journal}{\emph{arXiv preprint arXiv:1910.03771}}
  (\bibinfo{year}{2019}).
\newblock


\bibitem[Xiao et~al\mbox{.}(2023)]%
        {xiao2023smoothquant}
\bibfield{author}{\bibinfo{person}{Guangxuan Xiao} {et~al\mbox{.}}}
  \bibinfo{year}{2023}\natexlab{}.
\newblock \showarticletitle{Smoothquant: Accurate and efficient post-training
  quantization for large language models}. In \bibinfo{booktitle}{\emph{ICML}}.
  \bibinfo{pages}{38087--38099}.
\newblock


\bibitem[Xu et~al\mbox{.}(2022)]%
        {xu2022dense}
\bibfield{author}{\bibinfo{person}{Runxin Xu} {et~al\mbox{.}}}
  \bibinfo{year}{2022}\natexlab{}.
\newblock \showarticletitle{From dense to sparse: Contrastive pruning for
  better pre-trained language model compression}. In
  \bibinfo{booktitle}{\emph{AAAI}}, Vol.~\bibinfo{volume}{36}.
  \bibinfo{pages}{11547--11555}.
\newblock


\bibitem[Yao et~al\mbox{.}(2023)]%
        {yao2023comprehensive}
\bibfield{author}{\bibinfo{person}{Zhewei Yao} {et~al\mbox{.}}}
  \bibinfo{year}{2023}\natexlab{}.
\newblock \showarticletitle{A comprehensive study on post-training quantization
  for large language models}.
\newblock \bibinfo{journal}{\emph{arXiv preprint arXiv:2303.08302}}
  (\bibinfo{year}{2023}).
\newblock


\bibitem[Zhang et~al\mbox{.}(2022)]%
        {zhang2022opt}
\bibfield{author}{\bibinfo{person}{Susan Zhang} {et~al\mbox{.}}}
  \bibinfo{year}{2022}\natexlab{}.
\newblock \bibinfo{title}{OPT: Open Pre-trained Transformer Language Models}.
\newblock
\newblock
\showeprint[arxiv]{2205.01068}


\end{thebibliography}

\end{document}